\documentclass{PoS}

\title{Transverse Force on Quarks in DIS}

\ShortTitle{Transverse Force on Quarks in DIS}

\author{\speaker{Matthias Burkardt}\thanks{This work was supported by the DOE under grant number 
DE-FG03-95ER40965.}\\
        New Mexico State University\\
        E-mail: \email{burkardt@nmsu.edu}}

\abstract{Generalized Parton Distributions (GPDs) provide information on the distribution of quarks in impact paarmeter space. For transversely polarized nucleons, these impact parameter distributions are transversely distorted and this deviation from axial symmetry leads on average to a net transverse force from the spectators on the active quark in a DIS experiment. %The strength of that force can be related to twist-3 PDFs. 
This force when acting along the whole trajectory of the active quark leads to a transverse  single-spin asymmetries. For a longitudinally polarized nucleon target, 
the transverse force implies a torque acting on the quark
Orbital Angular Momentum (OAM). The resulting change in OAM as the quark leaves the target equals the difference between the Jaffe-Manohar and Ji OAMs.
}

\FullConference{XXIII International Workshop on Deep-Inelastic Scattering\\
		27 April - May 1 2015\\
		Dallas, Texas}

\newcommand{\be}{\begin{eqnarray}}
\newcommand{\ee}{\end{eqnarray}}
\newcommand{\bea}{\begin{eqnarray}}
\newcommand{\eea}{\end{eqnarray}}

\begin{document}

\section{Angular Momentum Decompositions}

Since the famous EMC experiments revealed that only a small fraction
of the nucleon spin is due to quark spins\cite{EMC}, 
there has been a great
interest in `solving the spin puzzle', i.e. in decomposing the
nucleon spin into contributions from quark/gluon spin and
orbital degrees of freedom.
In this effort, the Ji decomposition\cite{JiPRL}
\begin{equation}
\frac{1}{2}=\frac{1}{2}\sum_q\Delta q + \sum_q { L}_q^z+
J_g^z
\label{eq:JJi}
\end{equation}
appears to be very useful: through GPDs,
not only the quark spin contributions $\Delta q$ but also
the quark total angular momenta $J_q \equiv \frac{1}{2}\Delta q + 
{ L}_q^z$ (and by subtracting the spin piece also the
the quark orbital angular momenta $L_q^z$) entering this decomposition can be accessed experimentally. In the Ji decomposition (\ref{eq:JJi}) the quark OAM
is defined as the expectation value
\begin{equation}
{ L}_q^z= \int d^3r \langle PS| q^\dagger \left({\vec r} \times \frac{1}{i}{\vec D}
\right)^zq |PS\rangle /\langle PS|PS\rangle
\label{M012}
\end{equation}
in a nucleon state polarized in the $+\hat{z}$ direction. Here
${\vec D}={\vec \partial}-ig{\vec A}$ is the gauge-covariant
derivative.
The main advantages of this decomposition are that each term can be 
expressed as the
expectation value of a manifestly gauge invariant
local operator and that the
quark total angular momentum $J^q=\frac{1}{2}\Delta q+L^q$
can be related to GPDs\cite{JiPRL} 
and is thus accessible in deeply virtual Compton scattering and
deeply virtual meson production and can also be
calculated in lattice gauge theory. 

Jaffe and Manohar have proposed an alternative decomposition of the
nucleon spin, which does have a partonic interpretation, and in which also two terms, 
$\frac{1}{2}\Delta q$ and $\Delta G$,
are experimentally accessible \cite{JM}
\begin{equation}
\frac{1}{2}=\frac{1}{2}\sum_q\Delta q + \sum_q {\cal L}^q+
\Delta G + {\cal L}^g.
\label{eq:JJM}
\end{equation}
In this decomposition the quark OAM is defined as 
\begin{equation}
{\cal L}^q \equiv \int d^3r \langle PS|q^\dagger_+\!\left({\vec r}\times \frac{1}{i}{\vec \partial}
\right)^z \!\!q_+  |PS\rangle / \langle PS|PS\rangle,
\label{M+12}
\end{equation}
where light-cone gauge $A^+=0$ is implied. Although Eq. (\ref{M+12}) is not manifestly gauge invariant as written, gauge invariant extensions can be defined
\cite{lorce,hatta}. Indeed, manifestly gauge invariant definitions exist for each of the terms
in Eq. (\ref{eq:JJM}) which with the exception of $\Delta q$ involve matrix elements of
nonlocal operators. In light-cone gauge those nonlocal operators reduce to a local 
operator, such as Eq. (\ref{M+12}).

\section{TMDs and OAM from Wigner Distributions}

Wigner distributions can be defined as 
off forward matrix elements of non-local
correlation functions\cite{wigner,jifeng,Metz} with $P^+=P^{+\prime}$, $P_\perp = -P_\perp^\prime = \frac{q_\perp}{2}$
\begin{eqnarray}\label{eq:wigner}
%& &\hspace*{-1cm}
\!\!\!\!\!\!\!W^{\cal U}\!(x,\!{\vec b}_\perp,\! {\vec k}_\perp)\!
\equiv \!\!\!%\\ & &
\int \!\!\frac{d^2{\vec q}_\perp}{(2\pi)^2}\!\!\int \!\!\frac{d^2\xi_\perp d\xi^-\!\!\!\!}{(2\pi)^3}
e^{-i{\vec q}_\perp \!\!\cdot {\vec b}_\perp}\!
e^{i(xP^+\xi^-\!\!-{\vec k}_\perp\!\!\cdot{\vec \xi}_\perp)}
\langle P^\prime S^\prime |
\bar{q}(0)\Gamma {\cal U}_{0\xi}q(\xi)|PS\rangle .
%\nonumber
\end{eqnarray}
Throughout this paper, we will chose ${\vec S}={\vec S}^\prime = \hat{\vec z}$. Furthermore, we will focus on the 'good' component by selecting $\Gamma=\gamma^+$.
To ensure manifest gauge invariance, a Wilson line gauge link 
${\cal U}_{0\xi}$ connecting the quark field operators at position $0$ and $\xi$ is included. The issue of choice of path
for the Wilson line will be addressed below. 

In terms  of  Wigner distributions,  TMDs and OAM can be defined %respectively 
as \cite{lorce}
\begin{eqnarray}
%\langle {\vec k}_\perp\rangle_{\cal U} &=&
f(x,{\vec k}_\perp) &=& \int dx d^2{\vec b}_\perp d^2{\vec k}_\perp {\vec k}_\perp 
W^{\cal U}(x,{\vec b}_\perp,{\vec k}_\perp)\\
L_{\cal U}&=& \int dx d^2{\vec b}_\perp d^2{\vec k}_\perp \left({\vec b}_\perp \times {\vec k}_\perp \right)^z
W^{\cal U}(x,{\vec b}_\perp,{\vec k}_\perp).
\nonumber
\end{eqnarray}
No issues with the Heisenberg uncertainty principle arise here since only perpendicular combinations of position ${\vec b}_\perp$ and momentum ${\vec k}_\perp$ are
needed simultaneously in order to evaluate the integral for
$L_{\cal U}$.

A straight line connecting $0$ and $\xi$ for the Wilson line in ${\cal U}_{0\xi}$ results in
\cite{jifeng}
\begin{eqnarray}
L^q_{straight}
&=&
%\frac{ \int d^3{\vec r}\langle PS | 
%q^\dagger({\vec r}) \left( {\vec r}\times \frac{1}{i}{\vec D}\right)q({\vec r})^z|PS\rangle}
%{\langle PS |PS\rangle}=
L^q_{Ji}.
\label{eq:LJi}
\end{eqnarray}
However, depending on the context, other choices for the path in the Wilson link ${\cal U}$ should be made. Indeed for TMDs probed in SIDIS the path should be taken to be a straight line to $x^-=\infty$
along (or, for regularization purposes, very close to) the light-cone. This particular choice ensures proper inclusion of the FSI experienced by the struck quark as it leaves the nucleon
along a nearly light-like trajectory in the Bjorken limit. However, a Wilson line to
$\xi^-=\infty$, for fixed ${\vec \xi}_\perp$ is not yet sufficient to render Wigner distributions
manifestly gauge invariant, but a link at $\xi^-=\infty$ must be included to ensure manifest
gauge invariance. While the latter may be unimportant in some gauges, it is crucial in
light-cone gauge for the description of TMDs relevant for SIDIS \cite{jifengTMD}. 

\begin{figure}
\unitlength1.cm
\begin{picture}(10,2.2)(0.2,19)
\includegraphics{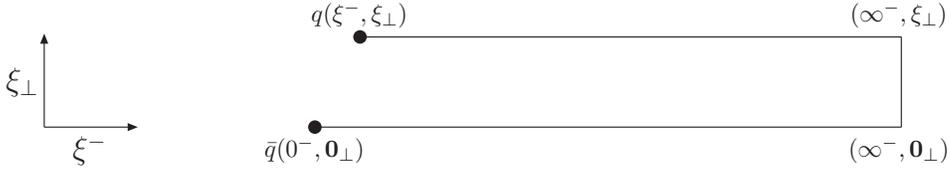}
\end{picture}
\caption{Illustration of the path for the Wilson line gauge link ${\cal U}^{+LC}_{0\xi}$ entering  $W^{+LC}$ }
%(\ref{eq:wigner}).}
\label{fig:staple}
\end{figure}

Let ${\cal U}^{+LC}_{0\xi}$ be the Wilson path ordered exponential obtained by first taking
a Wilson line from $(0^-,{\vec 0}_\perp)$ to $(\infty,{\vec 0}_\perp)$, 
then to $(\infty,{\vec \xi}_\perp)$, and then to $(\xi^-,{\vec \xi}_\perp)$, with each segment being a straight line (Fig. \ref{fig:staple}) \cite{hatta}. 
The shape of the segment at $\infty$ is irrelevant as the gauge field is pure gauge there, but it is still necessary to include a connection at $\infty$ and for
simplicity we pick a straight line. Likewise, with a similar 'staple' to $-\infty$ we define the Wilson path ordered exponential ${\cal U}^{-LC}_{0\xi}$, and using those light-like 
gauge links\footnote{Subtleties in regularizing/renormalizing such objects are addressed in Ref. \cite{Collins:what}.}, we define
\begin{eqnarray}\label{eq:wignerpm}
%& &\hspace*{-.7cm}
\!\!\!W^{\pm LC}\!(x,\!{\vec b}_\perp,\! {\vec k}_\perp\!)\!
%& &
\equiv\!\!\! \int \!\!\!\frac{d^2{\vec q}_\perp}{(2\pi)^2}\!\!\!\int\!\!\! \frac{d^2\xi_\perp d\xi^-\!\!\!\!}{(2\pi)^3}
e^{-i{\vec q}_\perp\!\! \cdot {\vec b}_\perp}\!
e^{i(xP^+\xi^-\!\!-{\vec k}_\perp\!\!\cdot{\vec \xi}_\perp)}
\!\langle P^\prime\! S^\prime |
\bar{q}(0)\Gamma {\cal U}^{\pm LC}_{0\xi}\!\!q(\xi)|P\!S\rangle .
%\nonumber
\end{eqnarray}
This definition for $W^{+LC}$ the same as that in Ref. \cite{hatta} and similar to that of $W_{LC}$ in Ref. \cite{jifeng} (the link segment at $\xi^-=\infty$ was not included in the definition of $W_{LC}$). 

In light-cone gauge $A^+=0$, only the segment at $\xi^-=\pm \infty$ contributes and 
the OAM looks similar to the local manifestly gaguge invariant expression, except
\be
{\vec r}\times {\vec A}({\vec r}) \longrightarrow {\vec r}\times {\vec A}(r^-=\pm \infty, {\bf r}_\perp).
\ee
From  PT invariance one finds that ${\cal L}_+^q={\cal L}_-^q$ \cite{hatta}.
In the Bashinsky-Jaffe definition of OAM ${\cal L}_{BJ}^q$ \cite{BJ}, the vector potential in the gauge covariant derivative is replaced by
\be
\frac{\int_{-\infty}^\infty dx^- A_\perp (r^-,{\bf r}_\perp)}{\int_{-\infty}^\infty dx^-} = \frac{1}{2}\left[ { A}_\perp (r^-= \infty, {\bf r}_\perp)+
{ A}_\perp (r^-= \infty, {\bf r}_\perp)\right],
\ee
and is thus equivalent to the light-cone-staple definition
\be
{\cal L}_{BJ}^q = \frac{1}{2}\left({\cal L}_+^q+{\cal L}_-^q\right) ={\cal L}_+^q={\cal L}_-^q.
\ee
Imposing $A^+=0$ does not completely fix the gauge as one can still make $r^-$-independent gauge transformations.
If one fixes this residual gauge invariant by imposing anti-symmetric boundary conditions
$A_\perp(r^-=-\infty,{\bf r}_\perp)=-A_\perp(r^-=-\infty,{\bf r}_\perp)$ the vector potential
at $r^-=\pm \infty$ cancels out in ${\cal L}_+^q+{\cal L}_-^q$ and therefore, with the
understanding of anti-symmetric boundary conditions at $r^-=\pm \infty$ the Jaffe-Manohar OAM
becomes also identical to ${\cal L}^q_\pm$. 

This last observation is crucial for understanding the difference 
between the Ji vs. Jaffe-Manohar OAM, which in light-cone gauge\footnote{As $L^q$ involves a manifestly gauge invariant local operator, it can be evaluated in any gauge.}
involves only the replacement ${ A}_\perp^i({\vec r}) \longrightarrow {A}_\perp^i(r^-=\pm \infty, {\bf r}_\perp)$.
Using
\begin{eqnarray}
{A}^i_\perp (r^-\!\!=\infty,{\bf r}_\perp)-{ A}^i_\perp (r^-,{\bf r}_\perp)
=\!\!\int_{r^-}^\infty\!\!\!\!\! dz^-
\partial_- {A}^i_\perp (z^-,{\vec r}_\perp)%\nonumber\\
= \!\!\int_{r^-}^\infty\!\!\!\!\! dz^- G^{+i}(z^-,{\vec r}_\perp)
\label{eq:kp}
\end{eqnarray}
where $G^{+\perp}=\partial_-A^\perp$ is the gluon field strength tensor in $A^+=0$ gauge. Note that 
\begin{equation}
-\sqrt{2}gG^{+y}\equiv -gG^{0y}-gG^{zy} = g\left(E^y-B^x
\right)
=g\left({\vec E}+{\vec v}\times {\vec B}\right)^y
\end{equation}
yields the $\hat{y}$ component of the color Lorentz force acting on a particle that moves with the velocity of light in the $-\hat{z}$ direction (${\vec v}=(0,0,-1)$) --- which is the direction of the 
momentum transfer in DIS \cite{QS,mb:force}. Thus the difference between the Jaffe-Manohar and Ji\footnote{Here we replaced $\gamma^0\rightarrow\gamma^+$ in $L^q$ as discussed in Ref. \cite{BC}.} OAMs
\be
{\cal L}^q-L^q = -g \!\!\int \!\!d^3x\!\left\langle P\!,\!S\right|\!
\bar{q}({\vec x})\!\gamma^+\!\!
\left[{ {\vec x}\! \times\! \!
\int_{x^-}^\infty \!\!\!\!\!dr^- F^{+\perp}(r^-,{\bf x}_\perp)
}\right]^z\!\!\!\!
q({\vec x}) \!\left| P\!,\!S\right\rangle/ \langle PS|PS\rangle
\label{eq:torque}
\ee
has the semiclassical interpretation of the change in OAM due to the torque from the FSI as the quark leaves the target:\cite{mb:torque}
while $L^q$ represents the local and manifestly gauge invariant OAM of the
quark {\it before} it has been struck by the $\gamma^*$, ${\cal L}^q$ represents 
the gauge invariant OAM {\it after} it has left the nucleon and moved to $r^-=\infty$.

\section{Intuitive Picture for the Torque from Final State Interactions}
In order to estimate the effect from the final state interactions on the quark OAM we first consider the effect
on a positron moving through the magnetic dipole field of an electron, which is polarized in the $+\hat{z}$ direction. This should be the most simple analogy to a proton polarized in the $+\hat{z}$ direction 
because more quarks are polarized in the same direction as the nucleon spin and the color-electric force between the active quark and the spectators is attractive.
\begin{figure}
\unitlength1.cm
\begin{picture}(10,7.5)(0,13.2)
\includegraphics{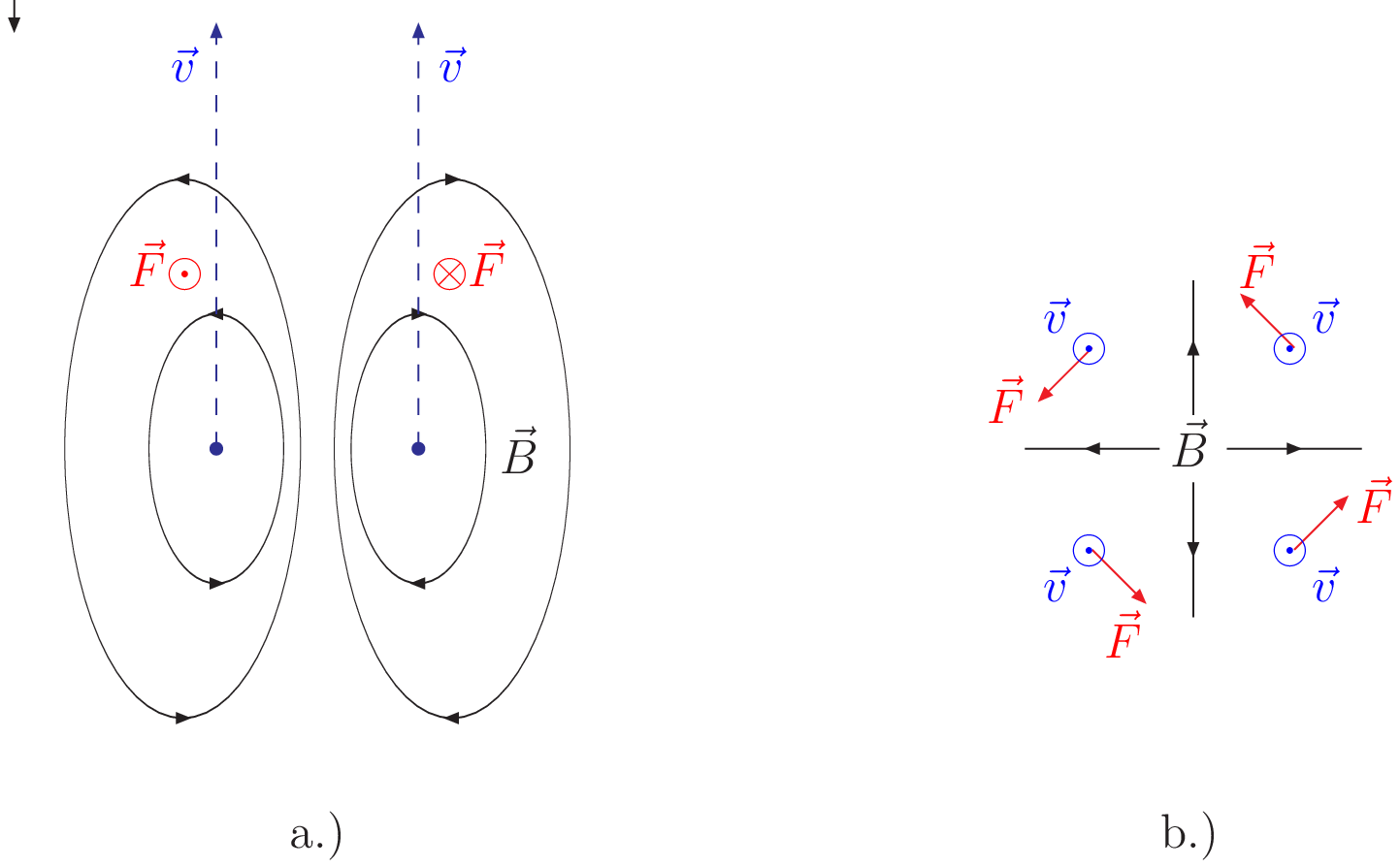}
\end{picture}
\caption{Illustration of the torque acting on a positron moving in the $-\hat{z}$ direction
through a magnetic dipole field caused by the magnetic moment of an electron polarized in the
$+\hat{z}$ direction. a.) side view; b.) top view. 
In this example the $\hat{z}$ component of the torque is negative as the positron leaves the
bound state.
}
\label{fig:dipole}
\end{figure}
As illustrated in Fig. \ref{fig:dipole} 
the magnetic FSI leads to a negative torque. Since the example was chosen such that the signs of polarization and forces are the same as in the nucleon this implies that the
color-magnetic torque acting on quarks ejected from the proton are negative as well.

The dipole example also illustrates why the torque vanishes immediately after the absorption of the virtual photon: when comparing the $\perp$ magnetix field (for fixed $\perp$ position) at positions at the 'front' and
'back' side of the nucleon are equal and opposite, and therefore the torque at the original position of the active quark averages to zero
\begin{equation}
\langle P,S| \bar{q}(0) \left[ xF^{+y}(0)-yF^{+x}(0)\right] q(0)|P,S\rangle=0.
\end{equation}
Note that this cancellation is a consequence of averaging over all possible initial positions of the active quark.
For a specific initial position the initial torque will in general be nonzero.
While we used here an intuitive picture to motivate this result, it can also be shown to be a rigorous consequence of PT invariance. 
%\end{document}

\end{document}